# Reconstruction of showers in the calorimeter during the first flight of the CREAM balloon experiment

P. S. Marrocchesi[d], H. S. Ahn[b], P. Allison[c], M. G. Bagliesi[d], J. J. Beatty [c], Bigongiari, G.[d], P. Boyle[e], A. Castellina[f], J. T. Childers[g], N. B. Conklin[h], S. Coutu[h], M. A. DuVernois[g], O. Ganel[b], J. H. Han[i], H. J. Hyun[i], J. A. Jeon[i], K. C. Kim[b], J. K. Lee[i], M. H. Lee[b], L. Lutz[b], P. Maestro[d], A. Malinine[b], S. Minnick[j], S. I. Mognet[h], S. W. Nam[i], S. Nutter[k], H. Park[l], I. H. Park[i], N. H. Park[i], E. S. Seo[a,b], R. Sina[b], S. Swordy[e], S. Wakely[e], J. Wu[b], J. Yang[i], Y. S. Yoon[a], R. Zei[d], S.-Y. Zinn[b]

*(a) Dept. of Physics, University of Maryland, College Park, MD 20742 USA*
*(b) Inst. for Phys. Sci. and Tech., University of Maryland, College Park, MD 20742 USA*
*(c) Dept. of Physics, Ohio State University, Columbus, Ohio 43210, USA*
*(d) Dept. of Physics, University of Siena and INFN, Via Roma 56, 53100 Siena, Italy*
*(e) Enrico Fermi Institute and Dept. of Physics, University of Chicago, Chicago, IL 60637, USA*
*(f) IFSI sez. di Torino and INFN, 4 Corso Fiume 4, 10133 Torino, Italy*
*(g) School of Physics and Astronomy, University of Minnesota, Minneapolis, MN 55455, USA*
*(h) Dept. of Physics, Penn State University, University Park, PA 16802, USA*
*(i) Dept. of Physics, Ewha Womans University, Seoul, 120-750, Republic of Korea*
*(j) Dept. of Physics, Kent State University Tuscarawas, New Philadelphia, OH 44663, USA*
*(k) Dept. of Physics and Geology, Northern Kentucky University, Highland Heights, KY 41099, USA*
*(l) Dept. of Physics, Kyungpook National University, Taegu, 702-701, Republic of Korea*
Presenter: Pier Simone Marrocchesi (marrocchesi@pi.infn.it), ita-marrochesi-P-abs1-he15-oral

The Cosmic Ray Energetics And Mass (CREAM) balloon-borne experiment was first flown from Antarctica in December 2004. The instrument includes a tungsten/Sci-Fi calorimeter preceded by a graphite target ($\sim 0.5$ interaction length and $\sim 1$ radiation length) where a hadronic shower is initiated by the inelastic interaction of the incoming nucleus. The fine granularity (1 cm) of the 20 radiation length calorimeter allows the imaging of the narrow electromagnetic core of the shower and the determination of the direction of the incident particle. Preliminary results, from the flight data, on the shower reconstruction capability of the instrument and on the observed shower properties are presented.

## 1. Introduction

CREAM is an experiment designed to perform direct measurements of the elemental composition and individual spectra of Very High Energy cosmic rays ($10^{12}$ to $10^{15}$ eV) in a series of balloon flights. During a record breaking flight of about 42 days, in December 2004 / January 2005, a total of $\sim$ 40 million triggers were collected.

The CREAM instrument features an excellent charge discrimination, in the range from proton to Fe and above, via multiple measurements of the particle charge with a pixelated silicon charge detector (SCD), a segmented timing-based particle-charge detector (TCD) and scintillating fiber hodoscopes. The energy measurement is carried out by two complementary techniques : a transition radiation detector (TRD) provides a measurement of the Lorentz factor for $Z \geq 3$ nuclei, while a sampling tungsten/scintillating fiber calorimeter, preceded by a graphite target, measures $Z \geq 1$ particles with almost energy independent resolution. A system of fiber hodoscopes provides dE/dx measurements and combined tracking capability with the TRD proportional tubes. The CREAM instrument layout and expected performances are described elsewhere [1, 2].

In this paper, we report a study, based on a preliminary analysis of the flight data, of the observed shower profiles in the calorimeter generated by the interactions of relativistic nuclei in the target.

## 2. The CREAM imaging calorimeter

The narrow electromagnetic core of a hadronic shower, initiated by the inelastic interaction of the primary nucleus in the graphite target, is imaged by a fine grained 20 radiation length ($X_0$) calorimeter ($50 \times 50$ cm$^2$), with active layers providing position measurements alternately in the X and Y directions. The calorimeter is longitudinally sampled every 1 $X_0$, while a lateral segmentation of 1 cm was chosen to match the Molière radius of the Tungsten absorber. The Tungsten/Sci-Fi stack is made of 20 tungsten plates with interleaved active layers instrumented with 1 cm wide ribbons of 0.5 mm diameter scintillating fibers. Details on the mechanical construction, beam test performance and energy calibration of the calorimeter can be found in [3]. The axis of the shower is reconstructed by the finely segmented imaging calorimeter, projected backwards and matched with the SCD pixels. The aim is to provide an adequate rejection of backscattering from the calorimeter and to allow the SCD to provide an unambiguous determination of the charge of the primary particle.

## 3. Longitudinal shower distribution

The longitudinal shower profiles generated in the calorimeter by ultra-relativistic nuclei were first simulated at a fixed particle energy (1 TeV) and at normal incidence (Fig.1) using the Fluka 2003.1b package [4]. Comparison with the expected longitudinal shower distribution for a proton of the same energy shows that the position of the shower maximum (in units of radiation length), for a nucleus of mass $A$, is lower than in the proton case by an amount proportional to $lnA$. This is expected if the longitudinal distributions scale, approximately, with the kinetic energy per nucleon.

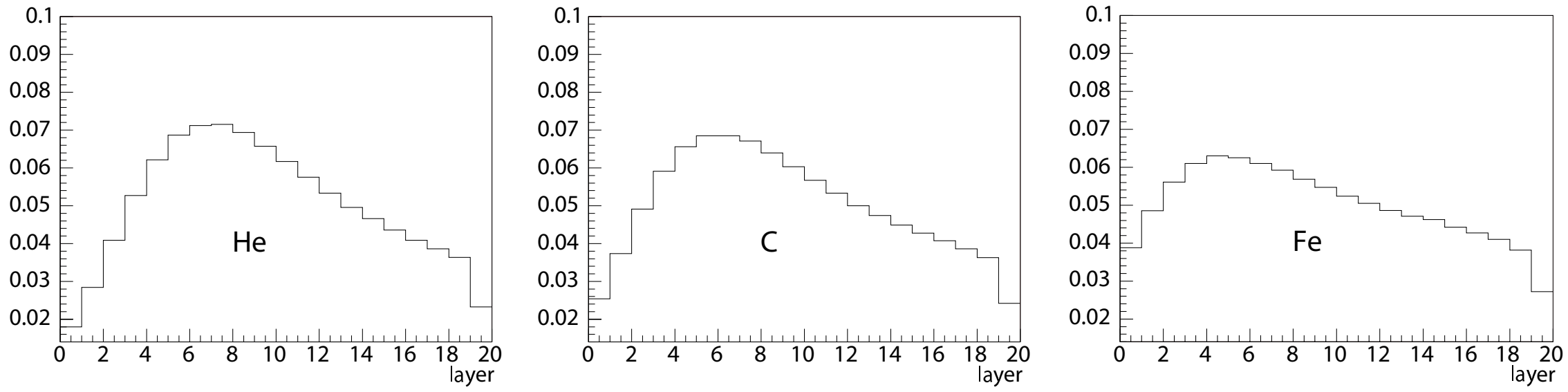

**Figure 1.** Monte Carlo simulation of the (normalized) average longitudinal shower development for a fixed particle energy of 1 TeV and normal incidence. From left to right: He, C and Fe nuclei.

The isotropic angular distribution and power law energy spectra expected for cosmic He and C nuclei incident within the CREAM acceptance were simulated using the same package for total particle energies between 2 and 6.5 TeV. It can be easily shown that a $lnA$ dependence of the shower maximum position is also expected in this case. Flight data were analyzed after application of preliminary calorimeter calibration, excluding channels which were not fully efficient during the flight (see Fig.2). A sample of tracks was selected with recontructed shower axis within the TCD acceptance and matched pixel in the SCD with a pulse height consistent with the dE/dx from a relativistic He or C nucleus. Using these events, the respective average longitudinal shower distributions were measured as shown in Fig.2, where the data points were found to be consistent with the Monte Carlo expectations (solid line).

A reduction of the average signal from the last scintillator plane is well visible both in the Monte Carlo simulations and in the data (Fig.1,2). This effect is due to the absence of an extra layer of absorber at the bottom of the calorimeter stack. The missing contribution to the signal in layer N is mainly due to the back-scattered

Compton component from absorber N+1, which becomes important when the shower energy drops below the threshold for pair-production.

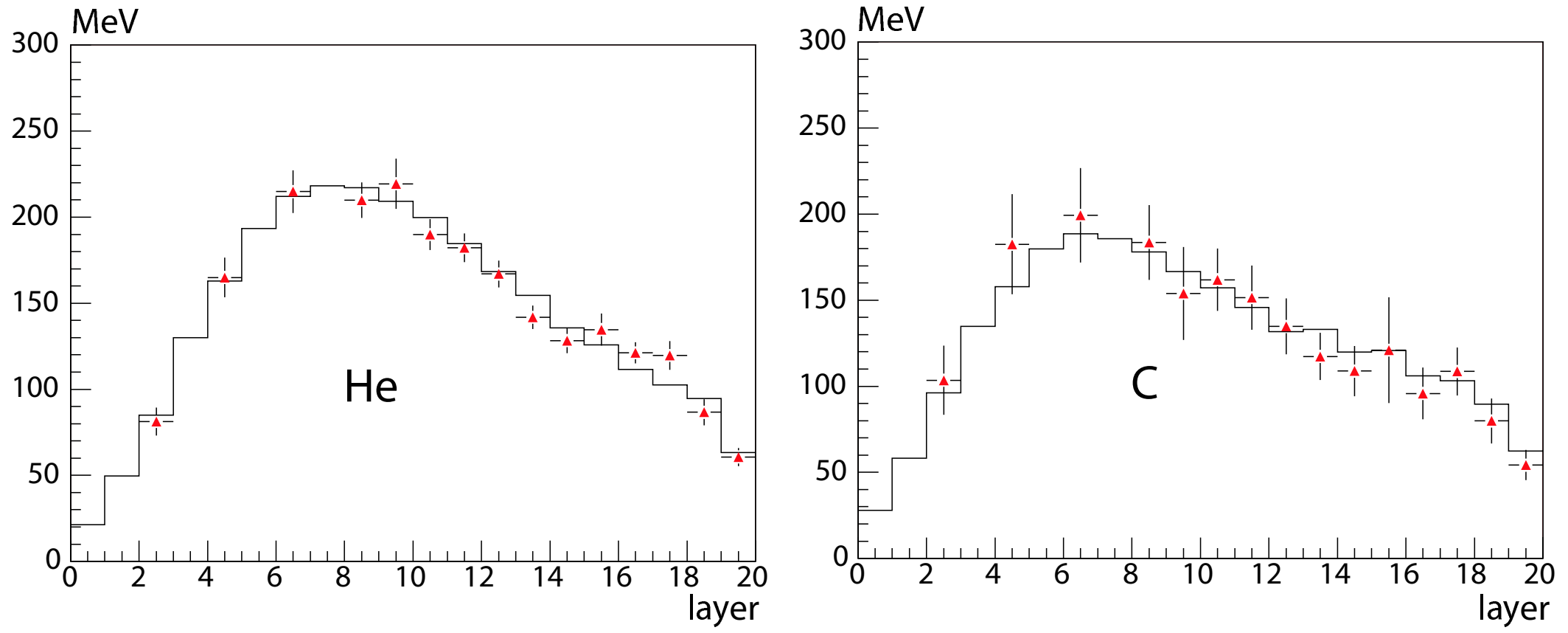

**Figure 2.** Average longitudinal shower development for relativistic He (left) and C (right) nuclei from flight data. Only statistical errors are shown. Layers 1, 2, 4, 6, 8 were excluded from the plot.

## 4. Integrated lateral shower distribution

The fine granularity of the calorimeter allows for the study of the lateral development of the shower. We define the integrated lateral shower profile $I_p(x)$, in a given layer $p$, as :

$$I_p(x) = \frac{\int_{x \geq 0}^{\infty} H(x)\, dx}{\int_{-\infty}^{\infty} H(x)\, dx}$$

where $H(x)$ is the pulse height of a ribbon at distance $x$ from the shower axis. By averaging $I_p(x)$ over all layers, the integrated average lateral profile $I(x)$ of the calorimeter is obtained. From the definition, it follows that a symmetrical lateral shower profile has $I(0) = 0.5$ and the distance $\tilde{x}$ corresponding to a given fraction $q$ of shower containement can be calculated by solving the equation $q(\tilde{x}) = 1 - 2I(\tilde{x})$.

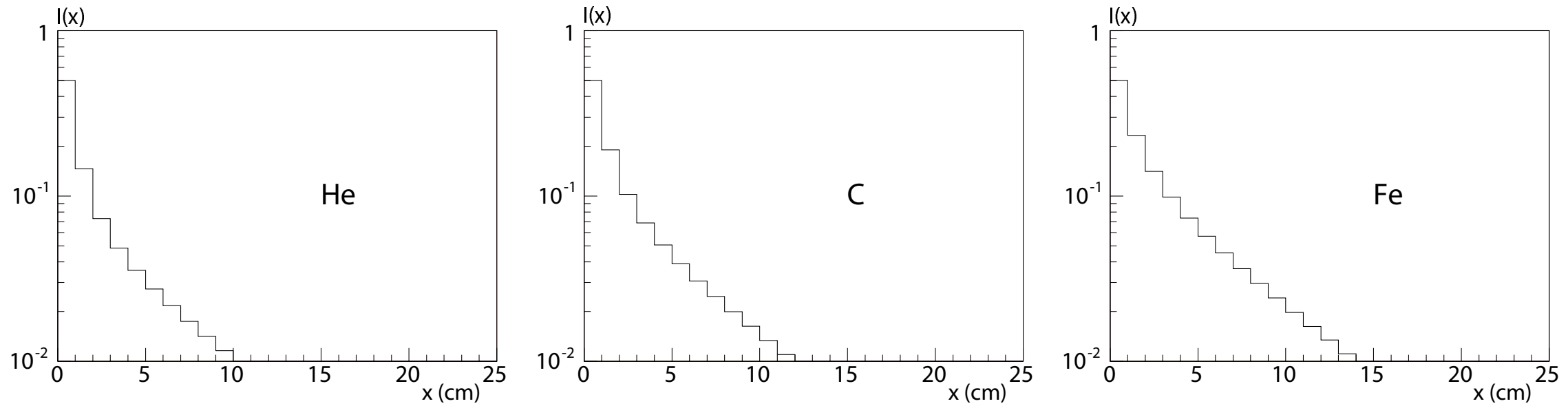

**Figure 3.** Monte Carlo simulation of the integrated average lateral shower development for a fixed particle energy of 1 TeV and normal incidence. From left to right: He, C and Fe nuclei.

The Monte Carlo predictions for He, C, Fe nuclei at a fixed particle energy of 1 TeV and normal incidence on the calorimeter are shown in Fig.3, where the shower width at 98% containment is found to increase for heavier mass nuclei. In the case of the flight data, the application of a sparsification threshold in the readout chain, introduces a truncation of the tails of the lateral distribution. Comparison between flight data and Monte Carlo (isotropic distribution, power law spectrum) is shown in Fig.4 using the same selection of He and C nuclei as in Sect.3. The measured points were found to be consistent with the simulation. The effect of the sparsification cut is also visible.

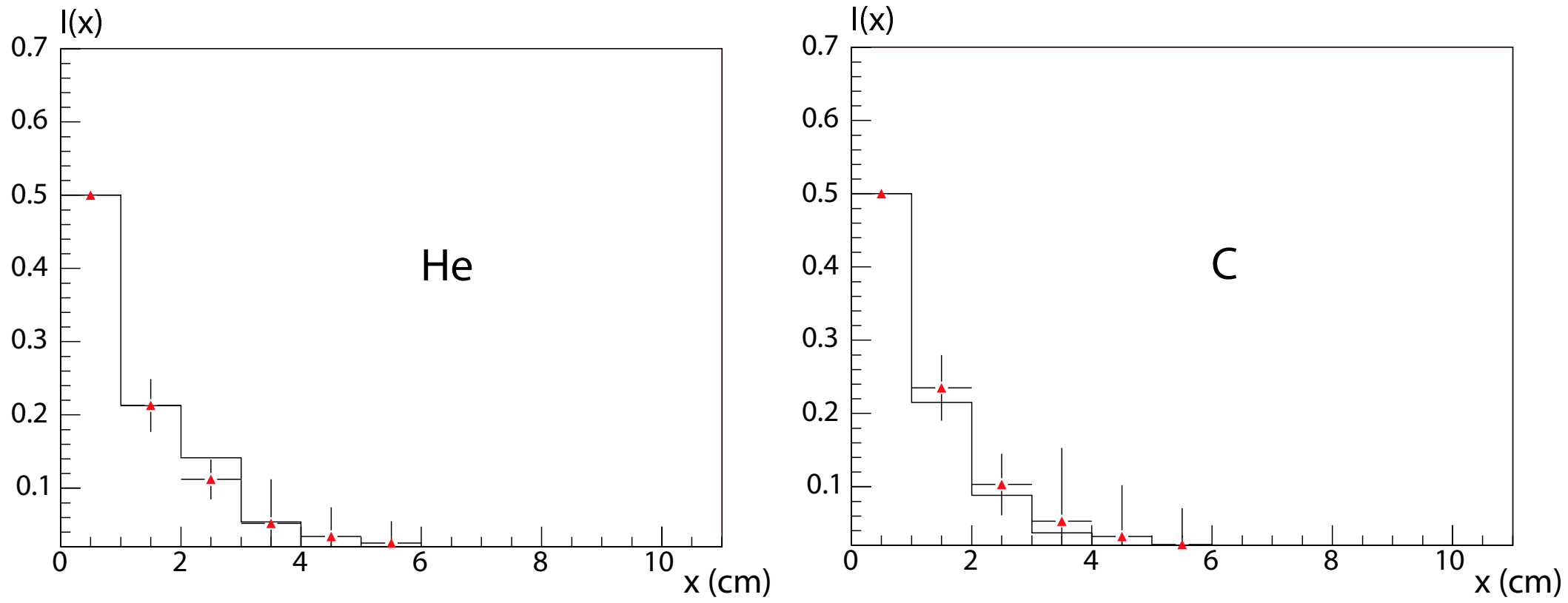

**Figure 4.** Integrated average lateral shower development for relativistic He (left) and C (right) nuclei from flight data. Only statistical errors are shown.

## 5. Conclusions

Showers originated by relativistic He and C nuclei in the CREAM calorimeter were found, at a preliminary stage of the analysis of the data from the first flight, to be consistent with Monte Carlo predictions.

## 6. Acknowledgements

This work was supported by NASA grants in the US and by INFN in Italy. We thank the National Scientific Balloon Facility and National Science Foundation for their support of the flight campaign and the Italian Antarctic Program (PRNA) for the support of the Italian participation in the flight operations in Antarctica.